\documentclass[12pt,a4paper,final]{iopart}

\usepackage[utf8]{inputenc} 

\usepackage{iopams}  
\usepackage{graphicx,calc}
\newlength\myheight
\newlength\mydepth
\settototalheight\myheight{Xygp}
\usepackage[breaklinks=true,colorlinks=true,linkcolor=blue,urlcolor=blue,citecolor=blue]{hyperref}

\newcommand*\inlinegraphics[1]{
	\settototalheight\myheight{Xygp}
	\settodepth\mydepth{Xygp}
	\raisebox{-\mydepth}{\includegraphics[height=\myheight]{#1}}
}

\usepackage{amsfonts}
\usepackage{gensymb}
\usepackage{booktabs}
\expandafter\let\csname equation*\endcsname\relax
\expandafter\let\csname endequation*\endcsname\relax
\usepackage{amsmath,bm,amssymb}
\usepackage{multirow}
\usepackage{enumitem}  
\usepackage{braket} 
\newcommand{\onlinecite}[1]{\hspace{-1 ex} \nocite{#1}\citenum{#1}} 

\usepackage[numbers,sort&compress]{natbib}

\usepackage[retainorgcmds]{IEEEtrantools}
\begin{document}

\title[Kim, Jeong, and Han]{Non-local Coulomb interaction and correlated electronic structure of TaS$_2$: A GW+EDMFT study}

\author{Taek Jung Kim, Min Yong Jeong, and Myung Joon Han}

\address{Department of Physics, Korea Advanced Institute of Science and Technology (KAIST), Daejeon 34141, Republic of Korea}
\ead{mj.han@kaist.ac.kr}

\vspace{10pt}


\begin{abstract}
	By means of $ab~initio$ computation schemes, we examine the low-energy electronic structure of monolayer TaS$_2$ in its low-temperature commensurate charge-density-wave structure. We estimate and take into account both local and non-local Coulomb correlations within cRPA (constrained random phase approximation) and GW+EDMFT (GW plus extended dynamical mean-field theory) method. Mott nature of its insulating phase is clearly identified. By increasing the level of nonlocal approximation from DMFT ($V=0$) to EDMFT and GW+EDMFT, a systematic change of charge screening effects is clearly observed while its quantitative effect on the electronic structure is small in the realistic Mott state.
\end{abstract}

\section{Introduction}
Understanding the effect of  Coulombic interaction between electrons in solid has long been a central theme of condensed matter physics {\cite{imada_metal-insulator_1998,lee_doping_2006}}. The relevant material systems are being extended to include various 2-dimensional (2D) van der Waals (vdW) materials {\cite{van_loon_competing_2018,zhang_emergence_2018,kim_mott_2019,kang_coherent_2020,van_loon_coulomb_2020,vano_artificial_2021}} and more recently their twisted combinations {\cite{cao_correlated_2018,cao_unconventional_2018,yankowitz_tuning_2019,burg_correlated_2019,liu_tunable_2020,shen_correlated_2020,cao_tunable_2020,wang_correlated_2020,xu_tunable_2021}}. The local onsite interaction (conventionally represented by $U$ in Hubbard model) can induce electron localization, magnetic moment and thereby leading to metal-to-insulator and magnetic phase transition as well as other related phenomena such as unconventional superconductivity {\cite{dagotto_correlated_1994,imada_metal-insulator_1998,dagotto_complexity_2005,lee_doping_2006,scalapino_common_2012}}. Non-local Coulomb interaction (denoted by $V$ in extended Hubbard model) also plays an important role. Typically, it enhances the itinerancy of electrons by screening Coulomb interactions or broadening the effective bandwidth {\cite{ayral_influence_2017,veld_bandwidth_2019}}. Out of its competition with the local correlation, intersite interactions can cause the instability toward the charge-ordered phase or charge density wave (CDW) {\cite{wigner_interaction_1934,wigner_effects_1938,lee_dynamics_1978,ayral_screening_2013,ayral_influence_2017,veld_bandwidth_2019}}. Recently, the effect of non-local Coulomb interaction receives increasing attentions in 2D material research \cite{rosner_two-dimensional_2016,raja_coulomb_2017,steinhoff_exciton_2017,steinke_coulomb-engineered_2020,van_loon_coulomb_2020}.

\begin{figure}[t]
	\centering
	\includegraphics[width=8.5cm]{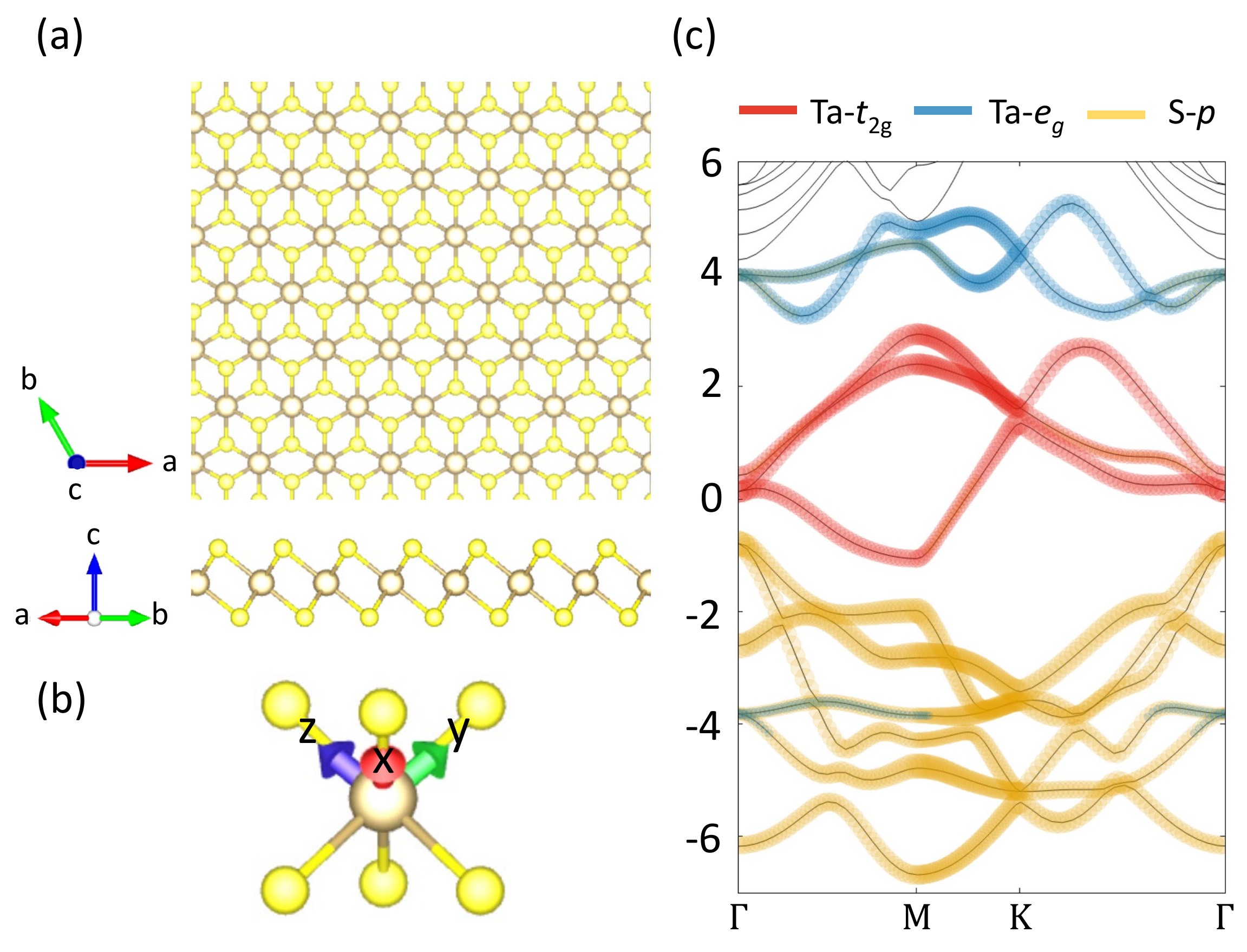}
	\caption{\label{fig.1} (a) Crystal structure of monolayer 1T-TaS$_2$. The top and side views are presented above and below respectively. (b) The local axis for octahedral structure of 1T-TaS$_2$ structure. (c) Electronic structure of monolayer 1T-TaS$_2$. Blue and red colors show Ta-t$_{2g}$, Ta-e$_{g}$ orbital contribution in the rotated local axis of the octahedral structure respectively, and yellow shows the p orbital contribution of S atoms. }
\end{figure}

\begin{figure}[t]
	\centering
	\includegraphics[width=8.0cm]{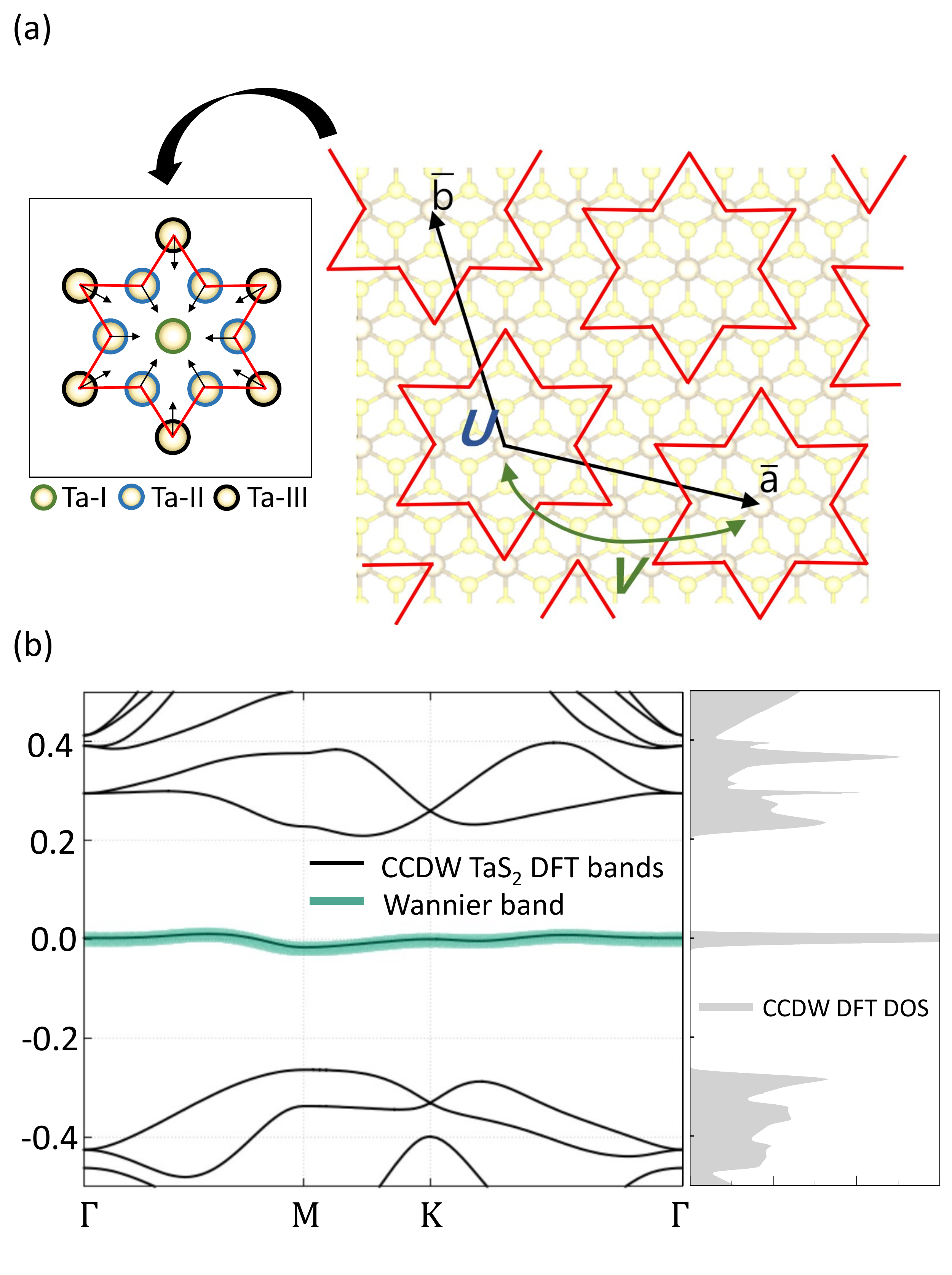}
	\caption{\label{fig.2} (a) Crystal structure of CCDW phase of TaS$_2$. Twelve Ta atoms move towards the central Ta constructing a unit of SOD and forming a single molecular orbital at the center. $U$ and $V$ refers to the onsite and the intersite Coulomb interaction of the molecular orbital, respectively. $\bar{a}$ and $\bar{b}$ are the unit cell vectors corresponding to CCDW phase. (b) The calculated DFT electronic structure for CCDW structure. }
\end{figure}

Among many correlated vdw materials, TaS$_2$ provides an intriguing case by displaying multiple quantum phases and their transitions { \cite{sipos_mott_2008,ang_atomistic_2015,yu_gate-tunable_2015,wang_band_2020}}. At high temperature, TaS$_2$ is known to be metallic and has 1T structure shown in Fig.~\ref{fig.1}(a). As temperature is lowered, it exhibits so-called `nearly commensurate CDW (NCCDW)', and then finally becomes insulating {\cite{wilson_charge-density_1975,fazekas_electrical_1979}}. Superconductivity is also observed below $T_C\approx$ 2--5~K by applying pressure, doping or electric field \cite{sipos_mott_2008,ang_atomistic_2015,yu_gate-tunable_2015}.  Importantly, the stabilization of the low temperature insulating phase is accompanied by commensurate CDW (CCDW) transition for which the long-range `star of David (SOD)' pattern of atomic rearrangement is well identified in both experiments and simulations (see Fig.~\ref{fig.2}(a))  \cite{wilson_charge-density_1975,fazekas_electrical_1979,giambattista_scanning_1990,liu_electron-phonon_2009,ge_first-principles_2010}. Therefore, unveiling the electronic property of CCDW SOD phase is important to understand the metal-insulator transition and the other related phases observed or suggested in this material at low temperature {\cite{sipos_mott_2008,ang_atomistic_2015,yu_gate-tunable_2015,wang_band_2020,law_1t-tas2_2017,ribak_gapless_2017,klanjsek_high-temperature_2017}}. Given that Ta atom has the 4+ formal valance, the SOD unit cell should have odd number of electrons ({\it i.e.}, $13 = 5d^{1} \times 13$), and CCDW-TaS$_2$ is expected to be metallic according to the conventional band theory. It is however in a sharp contrast to experiments \cite{lin_scanning_2020}. Several recent studies focusing  this issue suggested Mott mechanism or interlayer dimerization to be responsible for metal-to-insulator transition \cite{darancet_three-dimensional_2014,yu_electronic_2017,lee_origin_2019,petocchi_mott_2022, butler_mottness_2020,wang_band_2020,lee_distinguishing_2021,nicholson_modified_2022}. A crucially important challenge here is to estimate the strengths of Coulomb interactions \cite{fuhrmann_mott_2006}.

Also related is another intriguing possibility in this material, namely quantum spin liquid (QSL) phase. {Since the low-temperature CCDW phase makes a triangular lattice } and carries nominally S = 1/2 local moment, the low temperature TaS$_2$ was suggested as a QSL candidate \cite{law_1t-tas2_2017}. This idea is supported by the absence of long-range order down to very low temperatures and the gapless spin excitations \cite{law_1t-tas2_2017,ribak_gapless_2017,klanjsek_high-temperature_2017}. Notably, a recent study highlighted the effect of non-local Coulomb interaction on the magnetic nature related to QSL \cite{chen_controlling_2022}. However, most of previous computational studies on CCDW-TaS$_2$ have been limited to density functional theory (DFT), DFT+$U$ and single-site DMFT \cite{darancet_three-dimensional_2014,yu_electronic_2017,yi_coupling_2018,jiang_two-dimensional_2021}, which is the main motivation of our current study.

In this paper, we examine the electronic structure and the effect of correlations in monolayer CCDW-TaS$_2$. First, we try to directly estimate the interaction strengths in terms of both $U$ and $V$ based on random phase approximation (RPA) \cite{aryasetiawan_frequency-dependent_2004,sasioglu_effective_2011,nakamura_respack_2021}. The correlated electronic structure is calculated within GW+EDMFT (GW plus extended dynamical mean-field theory) \cite{sun_extended_2002,biermann_first-principles_2003,sun_many-body_2004,ayral_screening_2013} which enables us to deal non-perturbatively with local and non-local Coulomb interaction.  Our results show that monolayer TaS$_2$ is a Mott insulator mainly due to the local interaction. Even if the non-local Coulomb interaction is sizable, its effect on the electronic structure is quite small, largely suppressed in the Mott phase, and insufficient to achieve the charge-ordered phase.

\section{Computational details}

\subsection{DFT and extended Hubbard model construction}

We performed DFT calculations of monolayer 1T- and CCDW-TaS$_2$ with Vienna Ab initio Simulation Package (VASP) \cite{kresse_ab_1993,kresse_efficiency_1996}. The cell parameters and the internal coordinates were optimized with the force criterion of 0.1 meV/$\mathrm{\AA}$. The GGA-PBE functional \cite{perdew_generalized_1996} was used and the $25\mathrm{\AA}$ vacuum taken into account to simulate the monolayer. $21\times21\times1$ ($8\times8\times1$) k-grid and 500~eV (400~eV) cutoff energy were adopted for 1T-TaS$_2$ (CCDW-TaS$_2$).

To describe the effect of electronic correlations of low-energy state in CCDW-TaS$_2$, we consider the single-band extended Hubbard model:
\begin{equation}
\label{eq.1;Hubbard}
H=-\sum_{ij\sigma}t_{ij}c^{\dagger}_{i\sigma}c_{j\sigma}+U\sum_{i}n_{i\uparrow}n_{i\downarrow}+V\sum_{<ij>}n_{i}n_{j}-\mu\sum_{i}n_{i}
\end{equation}
where $c^{\dagger}_{i\sigma}$ and $c_{i\sigma}$ is the creation and annhilation operator of electron, respectively, with spin $\sigma=\{{\uparrow,\downarrow}\}$ at site $i$. $n_{i}=c^{\dagger}_{i\sigma}c_{i\sigma}$ is electron number operator at site $i$, and $\sum_{<ij>}$ denotes the nearest-neigbor summation. The hopping amplitude $t_{ij}$ between two site $i$ and $j$ was calculated with the maximally localized Wannier function (MLWF) method \cite{marzari_maximally_1997,souza_maximally_2001}. $U$ and $V$ is the local and non-local Coulomb interaction, respectively. The realistic estimation of these two interaction parameters is one of the main issues of current study. For this purpose, we used (c)RPA method as implemented in RESPACK code  \cite{nakamura_respack_2021}.

\subsection{GW+EDMFT}

The Hamiltonian is solved within GW+EDMFT whose standard self-consistent loop is briefly summarized below for the completeness of presentation \cite{ayral_screening_2013,ayral_influence_2017}.

\begingroup
\renewcommand\labelenumi{(\theenumi)}
\begin{enumerate}[label=(\roman*)]
	\item Calculate the lattice green function $G$ and the screened Coulomb interaction $W$:
	\begin{center}
		$G(\boldsymbol{k},i\omega_n)=[i\omega_n-t(\boldsymbol{k})+\mu-\Sigma(\boldsymbol{k},i\omega_n)]^{-1},$\\
		$W(\boldsymbol{q},i\Omega_m)=v(\boldsymbol{q})[1-v(\boldsymbol{q})P(\boldsymbol{q},i\Omega_m)]^{-1},$
		\label{item1:GandW}
	\end{center}
	where $\boldsymbol{k}$ and $\boldsymbol{q}$ are the crystal momentum vectors. $i\omega_n$ and $i\Omega_m$ refers to the fermionic and bosonic Matsubara frequency, respectively. The momentum space $t(\boldsymbol{k})$ corresponds to the Fourier transformed $t_{ij}$. $\Sigma(\boldsymbol{k},i\omega_n)$ and $P(\boldsymbol{q},i\Omega_m)$ represents the electron self-energy and the polarization function, respectively.  The bare Coulomb interaction $v(\boldsymbol{q})$ is given by $v(\boldsymbol{q})=2V[\mathrm{cos}(q_x)+\mathrm{cos}(\frac{1}{2}q_x+\frac{\sqrt{3}}{2}q_y)+\mathrm{cos}(-\frac{1}{2}q_x+\frac{\sqrt{3}}{2}q_y)]$. In practice, we start the first loop from $\Sigma(\boldsymbol{k},i\omega_n)=P(\boldsymbol{q},i\Omega_m)=0$.

	\item Compute the fermionic and bosonic Wiess fields $\mathcal{G}$ and $\mathcal{U}$:
	
	\begin{center}
		$\mathcal{G}(i\omega_n)=[G_\mathrm{loc}(i\omega_n)+\Sigma_\mathrm{loc}(i\omega_n)]^{-1},$\\
		$\mathcal{U}(i\Omega_m)=[W_\mathrm{loc}(i\omega_n)+P_\mathrm{loc}(i\Omega_m)]^{-1},$
		\label{item2:Wiess}
	\end{center}	
	where $A_\mathrm{loc}(i\omega_n)=\sum_{\boldsymbol{k}}A(\boldsymbol{k},i\omega_n)$ for any $A$.
	
	\item Solve the following impurity model to obtain the impurity self-energy $\Sigma_\mathrm{imp}(i\omega_n)$ and polarization $P_\mathrm{imp}(i\Omega_m)$:	
	\begin{center}
		$S_\mathrm{imp}=\int\int^{\beta}_{0}d\tau d\tau'\bar{c}(\tau)[-\mathcal{G}^{-1}(\tau-\tau')]c(\tau')+\frac{1}{2}\int\int^{\beta}_{0}d\tau d\tau'\mathcal{U}(\tau-\tau')n(\tau)n(\tau')$
		\label{item3:imp}
	\end{center}	
	where $A(\tau)$ is the fourier transormation of $A(i\omega_n)$, and $\bar{c}(\tau)$ and $c(\tau)$ are the Grassmann field for electrons.  Here we took the continuous-time hybridization expansion quantum Monte Carlo impurity solver as implemented in ComDMFT package \cite{choi_comdmft_2019}.

	\item Construct the new self-energy and polarization function from the calculated $\Sigma_{\rm imp}(i\omega_n)$ and $P_{\rm imp}(i\Omega_m)$:
	\begin{center}
		$\Sigma(\boldsymbol{k},i\omega_n)=\Sigma_\mathrm{imp}(i\omega_n)+\Sigma_\mathrm{nonloc}^\mathrm{GW}(\boldsymbol{k},i\omega_n),$\\
		$P(\boldsymbol{q},i\Omega_m)=P_\mathrm{imp}(i\Omega_m)+P_\mathrm{nonloc}^\mathrm{GW}(\boldsymbol{q},i\Omega_m),$\\
		\label{item4:GWembedding}
	\end{center}		
	where $\Sigma^\mathrm{GW}(\boldsymbol{k},i\omega_n)=-\sum_{\boldsymbol{q},i\Omega_m}G(\boldsymbol{k}+\boldsymbol{q},i\omega_n+i\Omega_m)W(\boldsymbol{q},i\Omega_m), P^\mathrm{GW}(\boldsymbol{q},i\Omega_m)=2\sum_{\boldsymbol{k},i\omega_n}G(\boldsymbol{k}+\boldsymbol{q},i\omega_n+i\Omega_m)G(\boldsymbol{k},i\omega_n),$
	and $A_\mathrm{nonloc}(\boldsymbol{k},i\omega_n)=A(\boldsymbol{k},i\omega_n)-A_\mathrm{loc}(i\omega_n)$.

	\item Go back to the step~\ref{item1:GandW} and repeat the calculations until $G(\boldsymbol{k},i\omega_n)$ and $W(\boldsymbol{q},i\Omega_m)$ are converged.

\end{enumerate}
\endgroup

\subsection{Coulomb interaction parameters}
\label{subsec:coulombint}

The direct first-principles calculation of interaction parameters corresponding to the low-energy CCDW molecular orbital is challenging due to the enlarged unitcell size (see Fig.~\ref{fig.2}). Therefore, previous studies of TaS$_2$ and NbSe$_2$ extracted the $U$ values from the undistorted 1T-structure \cite{kamil_electronic_2018,darancet_three-dimensional_2014}. Here we adopted the  same approach of these two papers. In the sense that our calculation is based on (c)RPA, it is very similar with Ref.~\onlinecite{kamil_electronic_2018}. In addition, we adopted two different approaches and also calculated the intersite interaction $V$ which has not been reported before.

In Ref.~\onlinecite{kamil_electronic_2018}, the $U$ value corresponding to CCDW-NbSe$_2$ was estimated as the average  of RPA-screened Coulomb interactions of Nb atoms in 1T-NbSe$_2$. We adopt it as our first way of calculating interaction parameter for CCDW-TaS$_2$: 
\begin{equation}
\label{eq.2;CoulombU}
U=\frac{1}{13^2}\sum_{\boldsymbol{R},\boldsymbol{R'}\in \inlinegraphics{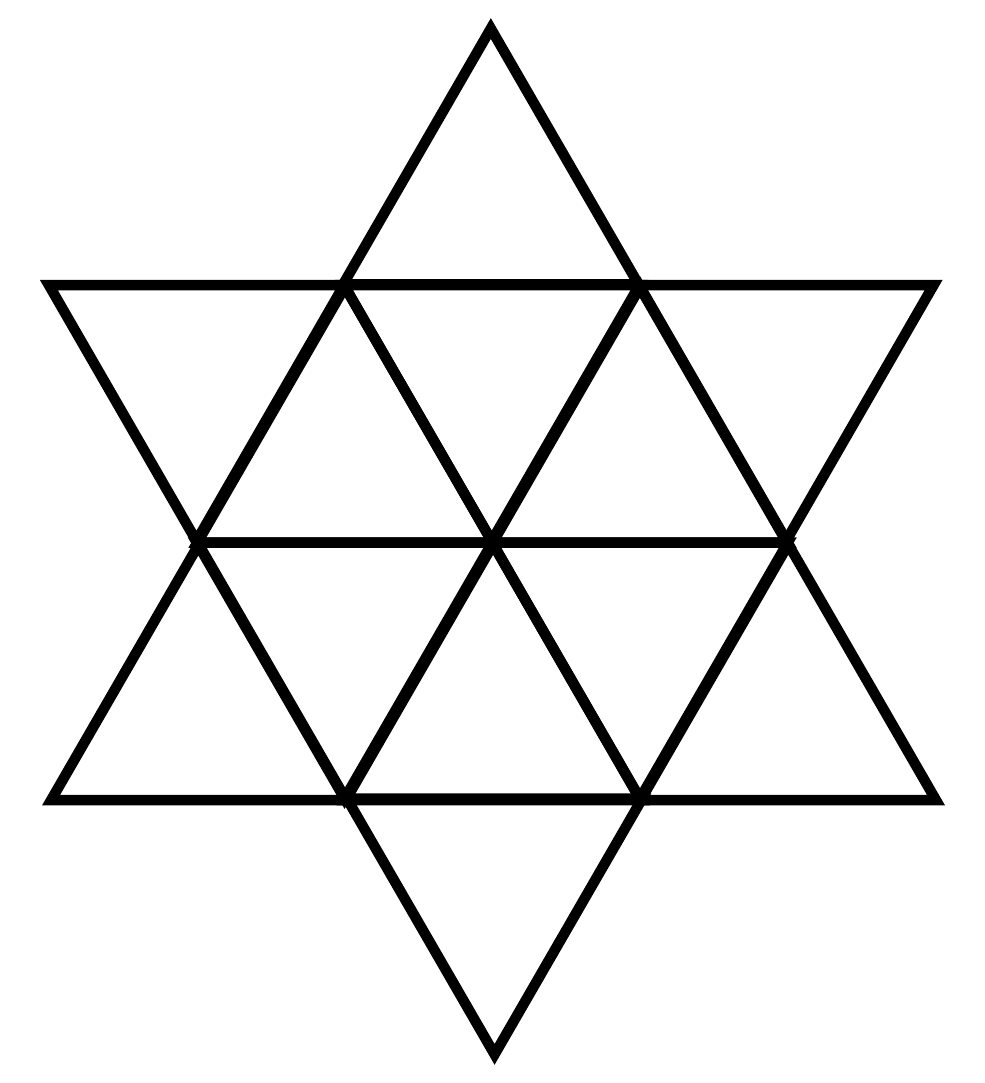}}U_{\boldsymbol{R}-\boldsymbol{R'}}
\end{equation}
where $U_{\boldsymbol{R}-\boldsymbol{R'}}$ is the Fourier transformed leading eigenvalue of RPA-screened Coulomb matrix $\mathcal{W}^{RPA}_{\boldsymbol{q}}$ represented in the eigenbasis of bare Coulomb matrix \cite{kamil_electronic_2018}. Here we calculated the screened Coulomb matrix for the $t_{2g}$ states, namely, the red-colored bands shown in Fig.~\ref{fig.1}(c). $\boldsymbol{R}$ and $\boldsymbol{R'}$ are the lattice cell vectors corresponding to 1T structure, and $\boldsymbol{q}$ is the lattice momentum. The white star symbol $\inlinegraphics{Figures/SOD1}$ represents the indices of Ta atoms in the same SOD (which is composed of 13 Ta atoms; see Fig.~\ref{fig.2}(a)).

We take this idea of Ref.~\onlinecite{kamil_electronic_2018} and extend it to compute the intersite interaction:
\begin{equation}
\label{eq.3;CoulombV}
V=\frac{1}{13^2}\sum_{\boldsymbol{R}\in \inlinegraphics{Figures/SOD1},\boldsymbol{R'}\in \inlinegraphics{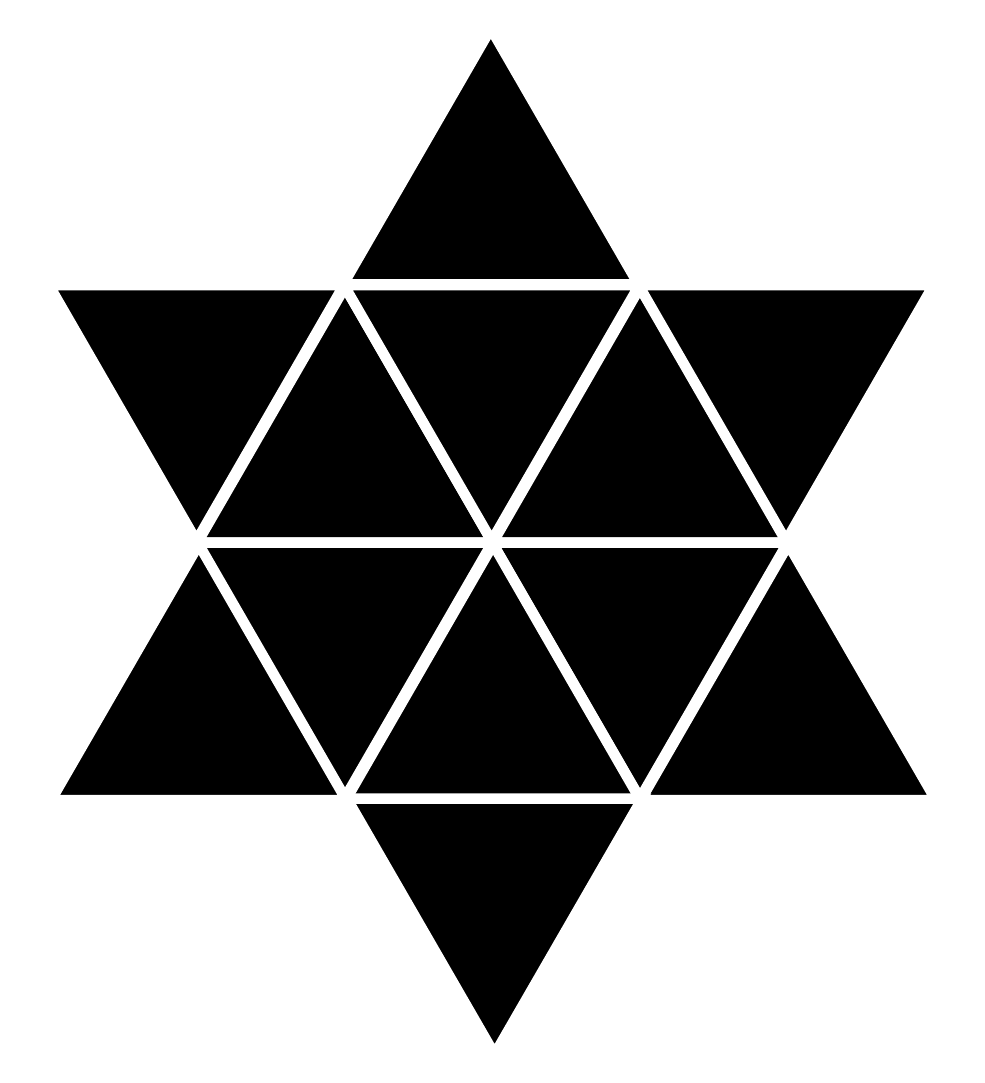}} U_{\boldsymbol{R}-\boldsymbol{R'}}.
\end{equation}
Here the black star symbol $\inlinegraphics{Figures/SOD2}$ refers to the indices of Ta atoms belonging to the nearest-neighboring SOD (indicated by unit cell vector  $\bar{a}$ or $\bar{b}$ ...; see Fig.~\ref{fig.2}(a)). The calculated $U$ and $V$ are presented in Table~\ref{tab.1} which will be referred to as `3 bands RPA' results hereafter.

Another possible way of calculating Coulomb parameters is to consider the weights of Ta atoms and to project them onto the molecular orbital of CCDW. In Ref.~\onlinecite{darancet_three-dimensional_2014}, $U$ for the molecular orbital was estimated from the atomic Ta-$d$ value, $\bar{U}^\mathrm{LR}$, obtained from the (so-called) linear response method suggested by Cococcioni et al.\cite{cococcioni_linear_2005}. The Coulomb interaction is estimated by calculating the projected weights of the Ta-$d$ orbitals ($\ket{d_{a}}$) with respect to CCDW molecular orbitals ($\ket{\Psi}$):
\begin{equation}
\label{eq.4;CoulombU_PRB}
\frac{U}{\bar{U}^\mathrm{LR}}=\sum_{a\in \inlinegraphics{Figures/SOD1}}|\braket{d_{a}|\Psi}|^4,
\end{equation}
where $a$ is the Ta atom index and Ta atoms are classified into three types, namely, Ta-I, Ta-II and Ta-III; see the inset of Fig.~\ref{fig.2}(a). Our calculations give rise to $|\braket{d_{a}|\Psi}|^2=$ 0.229, 0.064, and 0.035 for when Ta-I, Ta-II, and Ta-III, respectively, which are in good agreement with a previous study  \cite{darancet_three-dimensional_2014}.

Once again we extend this idea to calculate both $U$ and $V$:
\begin{equation}
\label{eq.5;CoulombU_PRB_my}
U=\sum_{\boldsymbol{R},\boldsymbol{R'}\in \inlinegraphics{Figures/SOD1}}|\braket{d_{\boldsymbol{R}}|\Psi}|^2 ~ |\braket{d_{\boldsymbol{R'}}|\Psi}|^2 ~\bar{U}_{\boldsymbol{R}-\boldsymbol{R'}},
\end{equation}
\begin{equation}
\label{eq.6;CoulombV_PRB_my2}
V=\sum_{\boldsymbol{R}\in \inlinegraphics{Figures/SOD1},\boldsymbol{R'}\in \inlinegraphics{Figures/SOD2}}|\braket{d_{\boldsymbol{R}}|\Psi}|^2 ~ |\braket{d_{\boldsymbol{R'}}|\Psi}|^2 ~\bar{U}_{\boldsymbol{R}-\boldsymbol{R'}}.
\end{equation}
The Coulomb interaction $\bar{U}_{\boldsymbol{R}-\boldsymbol{R'}}$ is caculated by cRPA method within 5-band model of 1T-TaS$_2$ ($t_{2g} + e_{g}$ bands; see Fig.~\ref{fig.1}(c)). The results are presented in Table~\ref{tab.1} which will be denoted as `5 bands cRPA'.

\begin{table}[t] 
	\caption{The calculated onsite ($U$) and intersite Coulomb interaction ($V$) for the molecular orbital of SOD charge density wave phase. The parameters are estimated from two different approaches, namely RPA and cRPA, as discussed in the main text.}
	\label{tab.1}
	\begin{center}
		\begin{tabular}{@{}ccccc@{}}
			\hline\toprule
			& \multicolumn{3}{c}{TaS$_2$}                                                                                                                                                                                                                 & NbSe$_2$                                                   \\  \cmidrule(l){1-1}\cmidrule(l){2-4} \cmidrule(l){5-5} 
			& \renewcommand{\arraystretch}{0.7} 
			\begin{tabular}[c]{@{}c@{}}3 bands\\ (RPA)\end{tabular} & \renewcommand{\arraystretch}{0.7}  \begin{tabular}[c]{@{}c@{}}5 bands\\ (cRPA)\end{tabular} &  \renewcommand{\arraystretch}{0.6}  \begin{tabular}[c]{@{}c@{}}Linear\\ response\end{tabular} & \renewcommand{\arraystretch}{0.7}  \begin{tabular}[c]{@{}c@{}}3bands\\ (RPA)\end{tabular} \\ 
			$U$ (eV) & 0.371                                                   & 0.653                                                                                                     & 0.180 \cite{darancet_three-dimensional_2014}                                                    & 0.300 \cite{kamil_electronic_2018}                                                   \\
			$V$ (eV) & 0.208                                                   & 0.359                                                                                                      & -                                                         & -                                                       
			\\ \bottomrule\hline
		\end{tabular}
	\end{center}
\end{table}

\section{Results and discussion}

\subsection{Non-interacting electronic structure}

Fig.~\ref{fig.1}(a) and (b) shows the crystal structure of 1T-TaS$_2$ and the local TaS$_6$ unit, respectively. The corresponding DFT-GGA band dispersion ($U=V=$ 0~eV) is  presented in Fig.~\ref{fig.1}(c) with the projected orbital characters at the local coordinates as defined in Fig.~\ref{fig.1}(b). The Ta-$t_{2g}$ states dominate the near Fermi level ($E_F$) region with $t_{2g}^{1}$ configuration. The higher energy bands above {+3~eV} and below {$-$1~eV} are mainly composed of Ta-$e_{g}$ and S-$p$ character, respectively. These overall electronic features are all in good agreement with previous studies \cite{mattheiss_band_1973,smith_band_1985,yan-bin_anisotropic_2007,yu_electronic_2017}.

By lowering temperature, 1T-TaS$_2$ is known to undergo CDW structural transitions. So-called nearly commensurate charge density wave (NCCDW) phase is stabilized below $T\approx$ 350~K and followed by CCDW at T$\approx$180~K \cite{wilson_charge-density_1975,fazekas_electrical_1979,giambattista_scanning_1990}. Figure~\ref{fig.2}(a) depicts the CCDW structure, so-called SOD pattern, in which twelve Ta atoms move toward the central one forming a 13-Ta-atom unit cluster {\cite{wilson_charge-density_1975,tosatti_nature_1976,fazekas_electrical_1979,brouwer_low-temperature_1980}}. In this phase, the band structure is also changed accordingly {\cite{perfetti_spectroscopic_2003,perfetti_time_2006,darancet_three-dimensional_2014,yu_electronic_2017}}. As shown in Fig.~\ref{fig.2}(b), a well-separated single electron band is solely responsible for the low energy region near $E_F$ while six bonding orbital states are fully occupied by twelve Ta-$d$ electrons well below $E_F$. It is noted that the near $E_F$ band is fairly flat and nondispersive which is therefore prone to metal-to-insulator transition when interactions come in to play {\cite{perfetti_spectroscopic_2003,perfetti_time_2006,darancet_three-dimensional_2014,yu_electronic_2017}}.


In order to perform GW+EDMFT calculations with both on-site ($U$) and inter-site ($V$) parameters (see the right pannel of Fig.~\ref{fig.2}(a)), we first construct the non-interacting Hamiltonian for CCDW-TaS$_2$ by means of MLWF method \cite{marzari_maximally_1997,souza_maximally_2001}. As shown in Fig.~\ref{fig.2}(b), this procedure well captures the target low energy electronic band; see the thick-green line (Wannier state) in good agreement with the black one (DFT band). The calculated hopping parameters of the first and the second neighbor hopping is 
$\left| t_1 \right| = 1.812$ and $\left| t_2 \right| = 0.814$ meV, respectively, and they are in good agreement with previous calculations \cite{pasquier_ab_2022,chen_controlling_2022}. Interestingly, the third neighbor hopping $\left| t_3 \right|=1.785$ meV is larger than $\left| t_2 \right|$ and comparable with $\left| t_1 \right|$ as recently reported also by Chen et al.\cite{chen_controlling_2022}.  

\subsection{Interaction parameters}

\begin{figure*}[t]
	\centering
	\includegraphics[width=16.5cm]{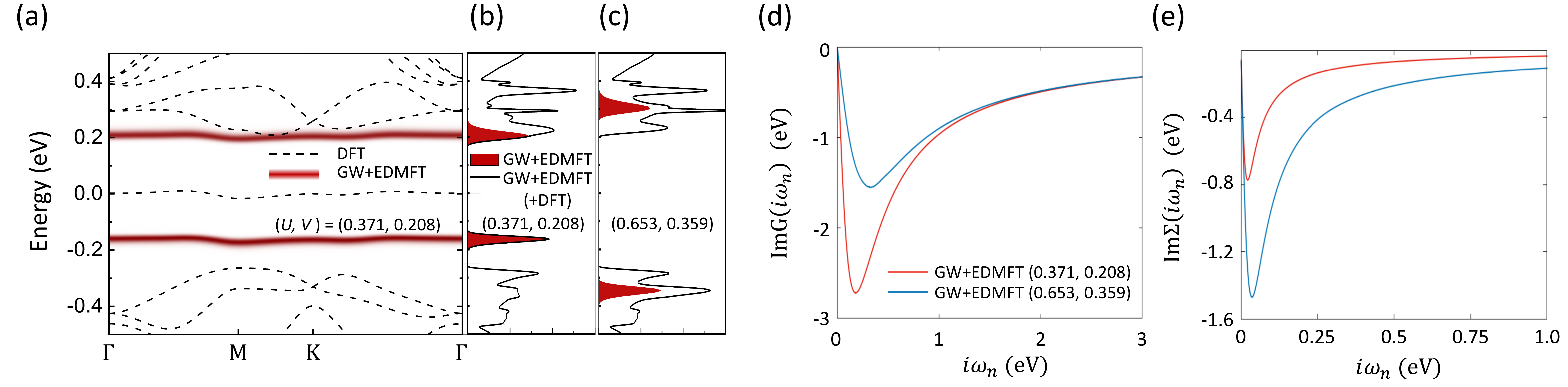}
	\caption{\label{fig.3} (a) The momentum-dependent spectral function of DFT (dashed lines) and GW+EDMFT (red line) calculated with Coulomb interactions estimated from 3 bands RPA model. The two numbers in parentheses indicate ($U$,$V$) in the unit of eV. (b-d) The calculated spectral function of GW+EDMFT with two different Coulomb interactions as estimated by (b) 3 bands RPA and (c) 5 bands cRPA model. The red colored data represent the density of states (DOS) of GW+EDMFT, and the black lines show the DFT DOS sum. (d) GW+EDMFT local self-energy computed by three different Coulomb parameter sets.  }
\end{figure*}

The metallic band structure of DFT-GGA demonstrates that Hubbard-type correlation is indispensable for reproducing insulating gap \cite{lin_scanning_2020}. In fact, previous studies show that both DFT+$U$ and single-site DMFT calculation give rise to the correct insulating state while the former also predicts the magnetic ground state \cite{darancet_three-dimensional_2014,yu_electronic_2017}. According to the Yu et al.'s DMFT calculation, Mott phase transition starts to occur at $U\sim$0.7~eV in bulk TaS$_2$ \cite{yu_electronic_2017}. {The effect of non-local interaction, on the other hand, has not been taken into account so far targeting the monolayer.}

The important first step to have the realistic correlated electronic structure is to estimate interaction strengths. For the T- or H-phase of transition metal dichalcogenides (TMDs), there have been several attempts to perform the direct estimatations \cite{huser_how_2013,ugeda_giant_2014,darancet_three-dimensional_2014,van_loon_competing_2018,kim_dynamical_2020,boix-constant_out--plane_2021,karbalaee_aghaee_ab_2022}. 
Due to the demanding computation cost, however, $ab~initio$ calculation of these parameters corresponding to SOD phase is challenging. Only available are a few indirect calculations just for on-site parameter $U$ \cite{darancet_three-dimensional_2014,kamil_electronic_2018} (see Table~\ref{tab.1}).

Here we, for the first time, directly calculate both $U$ and $V$ of CCDW-TaS$_2$ from first-principles method based on RPA/cRPA. As discussed in the previous Section, we adopted two different approaches and the results are summarized in Table~\ref{tab.1}. The calculated $U$ obtained from 3 bands RPA and 5 bands cRPA is 0.371 and 0.653~eV, respectively, both of which are notably larger than the result of a previous study based on linear response calculation \cite{darancet_three-dimensional_2014}.

It is interesting to compare the result of  Ref.~\onlinecite{darancet_three-dimensional_2014} with our 5 bands model both of which extract the interaction parameters  by taking the projected weights of Ta-$d$ orbitals onto CCDW-molecular orbitals. Interestingly, the on-site $U$ for Ta-$d$ (i.e., corresponding to the atomic on-site interaction) is larger in linear response calculation, $\bar{U}^{\mathrm{LR}}=2.27$~eV \cite{darancet_three-dimensional_2014}, than our cRPA $\bar{U}_{\boldsymbol{R}-\boldsymbol{R'}=\boldsymbol{0}}=1.83$~eV. Given that the final result of $U$ for CCDW phase is significantly larger in the 5 bands cRPA, {the difference is not attributed to the discrepancy between cRPA and linear response method. Rather,} it is clear from Eq.~(\ref{eq.5;CoulombU_PRB_my}) that our projection method not just takes all of Eq.~(\ref{eq.4;CoulombU_PRB}) portions into account but it also includes some inter-site contributions between different Ta atoms in a SOD molecular orbital.

Table~\ref{tab.1} also shows that the calculated $U$ for TaS$_2$ is larger than that for NbSe$_2$. As the computation method used by Kamil et al. \cite{kamil_electronic_2018} is based on 3 bands RPA, it is most straighforward to compare our result of $U$=0.371 eV (CCDW-TaS$_2$) with $U$=0.300 eV (CCDW-NbSe$_2$). According to a recent ARPES study by Nakata et al., the $U$ value is smaller for NbSe$_2$ than TaSe$_2$ \cite{nakata_robust_2021}. In addition, the value of TaSe$_2$ {may be} smaller than TaS$_2$ due to stronger screening by the chalcogen $p$-state { \cite{miyake_comparison_2010,karbalaee_aghaee_ab_2022}.}

It is not surprising and but understandable to have different values of interaction parameters depending on the computation methods and details. For example, the weaker correlation in 3 bands RPA model than 5 bands cRPA is naturally attributed to the all screening channels taken into account in the former (namely, it is not {\it constrained}). Among the available results, one possible way of choice is to take the value that is in the best quantitative agreement with a particular experimental measurement. For example, the gap size between the lower and upper Hubbard band has recently been measured by STM, and it gives rise to $U\sim$ 0.4~eV \cite{lin_scanning_2020}. Taking this experiment as the reference, our 3 bands model result provides the best quantitative agreement. For the sake of clarity, we will mainly present the result with this value in the remaining part of manuscript. However, it is obvious that all of our results coincidently predict the  large enough interaction strength with respect to the bandwidth $W\sim9|t|\sim0.02$~eV. 

\subsection{GW+EDMFT results}

Figure~\ref{fig.3}(a)--(c) summarize the GW+EDMFT results of spectral functions which were obtained by analytic continouation of Matsubara Green function using maximum entropy method \cite{jarrell_1996}. In sharp contrast to the metallic DFT band structure, GW+EDMFT result clearly shows the insulating gap. From the local spectral function shown in Fig.~\ref{fig.3}(b), the upper and lower Hubbard peak is clearly observed at about $\pm$0.2~eV which is in good agreement with a recent STM experiment of  monolayer TaS$_2$ \cite{lin_scanning_2020}. The result of 5 bands (cRPA) model is presented in Fig.~\ref{fig.3}(c). While this stronger correlation set of parameters also predicts the Mott insulting phase, the detailed electronic structure is noticeably different by positioning the Hubbard bands inside the uncorrelated states besides the enlarged gap size.

Figure~\ref{fig.3}(d) shows the imaginary part of local Green function calculated with two different parameter sets. It is clearly observed that Im$G(i\omega_n) $ goes to zero as $i\omega_n\rightarrow0$, confirming the zero spectral weight of the insulating phase at $E_F$. The minimum frequency point is lower when the 3 bands RPA set is used ($U$, $V$)$=$(0.371 eV, 0.208 eV) than the 5 bands cRPA, reflecting the smaller gap size. The calculated self-energy is presented in Fig.~\ref{fig.3}(e). In both sets of parameters, the steep slope of ${\partial{\rm Im}\Sigma}/{\partial\omega_n}$ at $i\omega_n\rightarrow0$ also clearly indicates Mott phase; $Z=[1-\frac{\partial{\rm Im}\Sigma}{\partial\omega_n}]^{-1} \approx 0$.



\begin{figure*}[t]
	\centering
	\includegraphics[width=15.0cm]{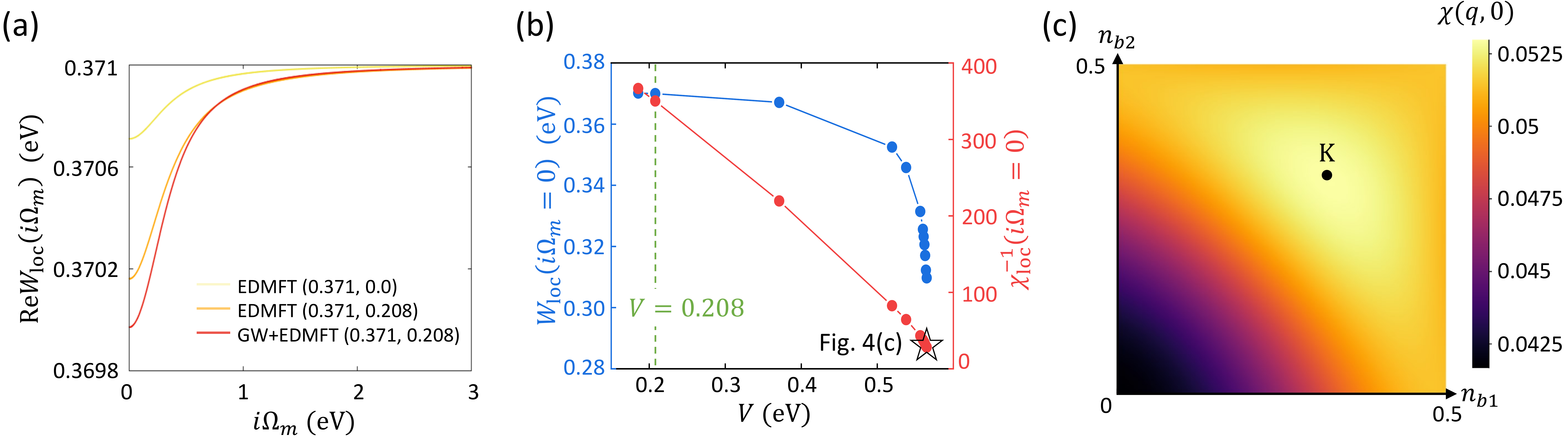}
	\caption{\label{fig.4} (a) The screened Coulomb interaction $W$ calculated by DMFT (yellow), EDMFT(orange), and GW+EDMFT(red) methods. (b) The static part of $W(i\Omega_m=0)$ (left y-axis) and $\chi^{-1}_{\mathrm {loc}}$ (right y-axis) as a function of $V$. Our estimation of $V$ value is indicated by vetical green-dotted line. The black star symbol indicates the value for Fig.~(c). (c) The momentum-dependent charge susceptibility $\chi(\boldsymbol{q},i\Omega_m=0)$ in the vicinity of charge-ordered phase; computed at $(U,V)=$(0.371~eV,0.565~eV)} 
\end{figure*}

Figure~\ref{fig.4}(a) shows the real part of screened Coulomb interaction. In order to examine the effect of non-local Coulomb interaction, we compare the three different cases: With the fixed value of onsite interaction $U=0.371~$eV, two different intersite interactions $V=$0 and 0.208 eV have been considered together with GW self-energy. {The frequency dependence shows that the effective Coulomb interaction approaches to the bare Coulomb $U=0.371$~eV  in the high-frequency limit, and decreases due to charge screenings in low frequency. It is noted that the low frequency part of $W_{\rm loc}$ gradually decreases as the more non-local interaction terms get involved.}  The same feature was also observed in the previous GW+EDMFT studies of the extended Hubbard model on a square lattice \cite{ayral_screening_2013,ayral_influence_2017}. Quantitatively, however, the screening by $V$ is quite small. Namely, the charge fluctuation is largely suppressed in this  Mott regime \cite{ayral_screening_2013,ayral_influence_2017}. In fact, the calculated $W(i\Omega_m=0)$ is close to the value of $U$=0.371 eV at around $V=0.208$ eV; see the blue points {on the green vertical dashed line} in Fig.~\ref{fig.4}(b).

On the other hand,  $W$ is significantly reduced in the large $V$ limit. It demonstrates that the charge fluctuations caused by $V$ further screen the Coulomb interactions \cite{ayral_screening_2013,ayral_influence_2017}. In this regime of {$V\gtrsim$} 0.5 eV, $W$ gets rapidly reduced, and simultaneously, the local charge susceptibility diverges {(see the red points)} {\cite{ayral_screening_2013,ayral_influence_2017}}: 
\begin{equation}
\chi_{\mathrm{loc}}^{-1}\rightarrow 0
\end{equation}
{where $\chi_{\mathrm{loc}}=\sum_{\boldsymbol{q}}\chi(\boldsymbol{q})$ and $\chi(\boldsymbol{q})=\Big(-P(\boldsymbol{q},i\Omega_m)[1-v(\boldsymbol{q})P(\boldsymbol{q},i\Omega_m)]^{-1}\Big)_{i\Omega_m=0}$.} In Fig.~\ref{fig.4}(c), we plot the momentum-dependent static charge susceptibility at $V$=0.565~eV (corresponding to the black star point in Fig.~\ref{fig.4}(b)). While the {diverging feature} at ($n_{b_{1}},n_{b_{2}}$)=(0,333,0.333) indicates the charge-ordering instability ($n_{b_i}$ is fractional coordinate of reciprocal lattice vector $b_{i}$), the largeness of $V$ implies that it is unlikely in this system {\cite{ayral_screening_2013,ayral_influence_2017}}.

When we were wraping up the current work, a closely related paper was published \cite{chen_causal_2022} in which Chen et al. conducted GW+EDMFT calculations of triangular lattice. With a main emphasis on the comparison of two different GW+EDMFT computation schemes, they also investigated TaS$_2$. A notable difference is that they did not try to estimate the interaction parameters as well as hopping integrals from first-principles, but instead focused on the comparison to the experiment.

\section{Summary}
{We investigated the electronic structure and the effect Coulomb interactions of monolayer TaS$_2$ with the combination of DFT, cRPA and GW+EDMFT. Both onsite and intersite correlation strengths are estimated based on two different $ab~initio$ cRPA approaches. The results coincidentally show that the monolayer CCDW-TaS2 is in the Mott insulator regime. The comparison of DMFT, EDMFT and GW+EDMFT exhibits a systematic change of the electronic structure as the level of nonlocal self-energy computations are varied while its overall effect is small in terms of electronic structure. Our work provides useful information and insight for understanding TaS$_2$, other TMDC materials and the related systems.}

\section{Acknowledgment}
{T.J.K thanks Sangkook Choi for useful discussion of code implementations of GW+EDMFT, and Siheon Ryee for the useful tips on the convergence issues of GW+EDMFT. This work was supported by the National Research Foundation of Korea (NRF) grant funded by the Korea government (MSIT) (Grant Nos.2021R1A2C1009303 and NRF-2018M3D1A1058754).}

\bibliography{TaS2_bib} 

\end{document}